\documentstyle[prd,aps,12pt]{revtex}

\begin{document}

\draft
\title{Photoproduction of the Eta-Prime Mesons as a New Tool
to Probe Baryon Resonances}
\author{J.-F. Zhang$^a$,
Nimai C. Mukhopadhyay$^{a}$
and M. Benmerrouche$^{b}$\\
$^a$ Physics Department\\ Rensselaer Polytechnic Institute\\
Troy, NY 12180-3590 \\
$^{b}$ Saskatchewan Accelerator Laboratory \\ University of Saskatchewan\\
Saskatoon, SK S7N 0W0}

\maketitle

\begin{abstract}

We examine eta prime
($\eta^{\prime}$) photoproduction, as a novel tool to study
baryon resonances around 2 GeV,  of particular interest to the
quark shell model, which predicts a number of them. We find important
r\^{o}les of  the  form factors at the strong vertices, and
 show that the
N$^*$(2080) can be probed efficiently by this reaction.
\end{abstract}



Thanks to the advent of the first continuous wave (cw) electron
machine in the 4-6 GeV region of electron energy, now in operation at
CEBAF, a powerful tool is at hand to probe the baryon resonances with
real and virtual photons. A  novel reaction to use in this context,
focus of this paper, is
\begin{equation}
\gamma +p\rightarrow\eta^{\prime}+p ,
\end{equation}
where $\eta^{\prime}$(958) is the heaviest member of the ground state
pseudoscalar meson nonet. Very little is known about this reaction
either theoretically or experimentally. Our paper aims at establishing the
importance of this reaction in probing the nucleon resonances around
2 GeV, whereabout the quark shell model predicts a rather rich
structure\cite{ref1}. The present experimental picture is very
confusing\cite{ref2}, not yet confirming many of these resonances,
  giving rise to what has been
termed a {\it missing resonances} problem\cite{ref3}. The threshold for the
reaction is $W=1896 MeV$, corresponding to a lab photon energy of
$E_{\gamma}=1447 MeV$. At present, the world supply of the
photoproduction data comes from the old work of the ABBHHM group at
DESY\cite{ref4}, consisting of just ten points of total cross-section over a
very broad energy range ($E_{\gamma}=$  1.7 to 5.2 GeV), with poor
energy resolution and statistics. New experiments, proposed\cite{ref5}
at CEBAF, would change this situation radically in the near future.

The eta prime meson is interesting in its own right, along with its close
relative, the $\eta(547)$, for a variety of reasons. First is the
question of the $\eta_1$, $\eta_8$ mixing angle. There are large discrepancies
between values obtained from the linear and quadratic mass formulas of
the nonet\cite{ref6}. There is also the  issue of the quark content of the
$\eta$ and $\eta^{\prime}$, on which some\cite{ref7} informations are
available from the $J/\psi\rightarrow $vector$+$pseudoscalar decays.
The question of the relevance of the SU(3)$\times$SU(3)
chiral perturbation theory\cite{ref8} is not clear here;
 no information is
available on its  application  to the reaction (1).
Finally, there is the old issue of the chiral U(1)
 symmetry breaking and the eta
mass\cite{ref9}, hence its possible relevance to the eta prime mass.

Of special interest to the reaction (1) are questions connected with
the effective Lagrangian of the $\eta^{\prime}NN$ interaction. We can
write this  at the tree level\cite{ref10,ref11}
\begin{equation}
L_{\eta^{\prime}NN}=g_{\eta^{\prime}}[-i\epsilon\bar{N}\gamma_{5}N\eta^{
\prime}+(1-\epsilon )(1/2M)\bar{N}\gamma_{\mu}\gamma_{5}N\partial^{\mu}
\eta^{\prime}],
\end{equation}
where $0\leq\epsilon\leq 1$, the coupling constant $g_{\eta^{\prime}}$
is essentially unknown. There is no compelling reason, from the heavy
mass of the $\eta^{\prime}$, to choose either $\epsilon =1$ (pseudoscalar(PS))
or $\epsilon =0$ (pseudovector (PV)) limit, or any particular value in
between. In this paper, we shall investigate the pseudoscalar coupling case
($\epsilon =1$).

We can make an estimate of the coupling constant $g_{\eta^{\prime}}$ by
using the quark-model mixing relation, where the singlet to octet mixing
angle is $\theta$. We have, with singlet and octet coupling taken as
$g_{\eta_{8}}$ and $g_{\eta_{1}}$ respectively,
\begin{equation}\left(
\begin{array}{c}
g_{\eta}\\g_{\eta^{\prime}} \end{array}\right) =\left( \begin{array}{cc}
\cos\theta & -\sin\theta \\
\sin\theta &\cos\theta\end{array}\right) \left(
\begin{array}{c} g_{\eta_{8}}\\g_{\eta_{1}}\end{array}\right) .
\end{equation}
We assume that  strange quark content of the
 nucleons is negligible and
take the $g_{\eta_{1}}/g_{\eta_{8}}\simeq \sqrt{2}$ \cite{ref12}.
This simply
follows from the $SU(3)$-flavor wavefunctions of the ${\eta_{8}}$
and ${\eta_{1}}$ configurations. From this, we can write
\begin{equation}
g_{\eta^{\prime}}=\frac{\sqrt{2}\cos \theta -\sin \theta}{\cos \theta +
\sqrt{2}\sin \theta} g_{\eta} .
\end{equation}
The coupling constant $g_{\eta}$ is also poorly known. But the $\eta$
photoproduction data have yielded\cite{ref10} a range,
$0.6\leq g_\eta^2/4\pi \leq 6.4,$ for $\epsilon =1$.
This allows us to vary the coupling of the eta prime meson to
the nucleon in the domain
\begin{equation}
1.9\leq g_{\eta^{\prime}}\leq 6.2,
\end{equation}
assuming $\theta \simeq 20^0$\cite{ref6},
 $\epsilon =1$. This would be the coupling
constant range we shall be  using in this work. We have found that the
process (1) is not very sensitive to this quantity.

The tree-level structure of the photoproduction amplitude can be
determined in parallel to that of the $\eta$ photoproduction\cite{ref10}.
 While there are obvious similarities, the
differences must be stressed, starting with  $g_{\eta^{\prime}}$ in (2).
The t-channel vector meson exchanges\cite{ref13}
 involve the $\rho$ and $\omega$
mesons, with  the product of their relevant
strong and electromagnetic coupling constants\cite{ref10}
constrained by the relations
\begin{eqnarray}
\lambda_\rho^{\prime} g_v^\rho+\lambda_\omega^{\prime}
 g_v^\omega&\simeq&4.1, \\
\lambda_\rho^{\prime} g_t^\rho+\lambda_\omega^{\prime}
 g_t^\omega&\simeq&12.1,
\end{eqnarray}
Relations (6) and (7) are obtained, in paralled to the case of the
$\eta$ meson\cite{ref10}, taking into account the $\eta^{\prime}$ structure in
the quark model.
 The experimentally measured ratio of the widths\cite{ref2}
of the radiative processes $\eta^{\prime}\rightarrow\rho^0 \gamma$ and
$\eta^{\prime}\rightarrow\omega\gamma$ is  about ten, while our quark model
estimate gives  11.7.

 The biggest difference from the eta photoproduction comes from the
intermediate $N^{*}$'s excited in the eta prime photoproduction. In the
eta photoproduction, $N^*$(1535) is known to be dominant\cite{ref10}.
 In the reaction (1)
specific relevance  of the $N^*$'s can be roughly determined from the work of
Capstick and Roberts(CR)\cite{ref1}. Based on  the strength  of the
product
\begin{equation}
\chi_{\lambda}=\Gamma^{1/2}_{\eta^{\prime}}A_{\lambda}/\Gamma_{0} ,
\end{equation}
where $A_{\lambda}$ is the electromagnetic excitation amplitude
$N\gamma\rightarrow N^*$ for helicity $\lambda$, $\Gamma_{\eta^{\prime}}$
and $\Gamma_{0}$ are the $\eta^{\prime}$-nucleon  and total decay widths
respectively, for a given $N^*$.
 We choose nine resonances in this work as candidates for
excitation in the reaction (1). These are: two S11 resonances ($N^*$(2030)
and $N^*$(2090)), three D13's ($N^*$(2055), $N^*$(2080) and $N^*$(2095)),
two D15's ($N^*$(2080) and $N^*$(2200)), one F15 ($N^*$(2000)) and one F17
($N^*$(1990)) respectively [here L2I2J are the quantum number of the
resonances in the meson-nucleon partial wave channel]. From our
analysis, we find the contributions\cite{ref14}
 of the spin-$5/2$ and spin-$7/2$ nucleon resonances
to be negligible. Hence we do not discuss these contributions in any detail.

The effective Lagrangians involving the $\gamma NR$ and $\eta^{\prime}NR$
vertices, where $R$ is a particular nucleon resonance, are needed
here\cite{ref10}.
 To illustrate this, we take the case of  the odd parity, spin-$3/2$,
$R$. The Lagrangian for the $\eta^{\prime}NR$ interaction is
\begin{eqnarray}
{\cal L}_{\eta^{\prime} NR}&=&\frac{f_{\eta^{\prime} NR}}{\mu} \bar{R}^\mu
\theta_{\mu\nu}(Z)\gamma_5 N \partial^\nu \eta^{\prime} + H.c.,
\end{eqnarray}
where $R^\mu$ is the  vector-spinor for $R$ and the tensor $\theta_{\mu\nu}$
is
\begin{equation}
\theta_{\mu \nu}(Z)=g_{\mu
\nu}+[\frac{1}{2}(1+4Z)A+Z]\gamma_\mu\gamma_\nu .
\end{equation}
We choose the point-transformation parameter $A$ to be $-1$ without any
loss of generality, and fit the parameter $Z$ and two other similar ones from
the electromagnetic vertices.

An interesting   feature of the reaction (1)
 is the r\^{o}le of the s- and u-dependence of the
form factor at the $\eta^{\prime}NR$ vertex.
 Not much
is known\cite{ref15}
 about this form factor either theoretically or experimentally.
 Fig. 1
demonstrates the importance of the use of a form factor, without which the
cross-section would  simply grow unphysically with energy. We discuss
below our choice of the form factors.

  The $s-$ and $u$-channel resonance excitation amplitudes are separately
gauge-invariant\cite{ref10}. Therefore, the choice of
form factors for these contributions
is relatively simple, but theoretically not rigidly fixed. The phenomenological
success in reproducing the shape of the experimental cross-section is the
important guide here. Thus, we utilize a form factor
for the s-channel resonance excitation\cite{ref16}
\begin{equation}
F(s)=1/[1+\frac{(s-M^2_{R}  )^2}{\Lambda^4}] .
\end{equation}
Here we use $\Lambda^2 =1.2 GeV^2$ for
the S11 resonances and $0.8 GeV^2$ for the D13 resonances. The values of
$\Lambda$'s are determined from the best fit.  A form similar to
(11), with u replacing s, would also do for the u-channel. For the
t-channel vector meson exchange, the form factor that we use
is standard\cite{ref17}:
\begin{equation}
F(t)=\left (\frac{\Lambda^2 -M_{V}^2}{\Lambda^2 -t}\right )^{2}
\end{equation}
where we take
 $\Lambda^2 =1.2 GeV^2$, and $M_{V}$ is the mass of the vector meson
exchanged.

For the nucleon Born terms one should also attach form factors at the
strong vertices, since the intermediate nucleon is off-shell. This,
in general, will not preserve gauge invariance and care is needed to
maintain gauge invariance. We use F(s) in the form (11), with M
replacing $M_R$, consistent with the
requirement of gauge invariance, choosing $\Lambda^2 =1.2 GeV$.

Fig.2  shows the main result of this paper. The available total cross-section
data from the DESY experiment are well-described by our effective
Lagrangian approach. {\it The main contribution to the photoproduction
amplitude comes from the $N^*$(2080)  alone.} In this sense, the
eta prime photoproduction reaction is helpful to illuminate the
property of this resonance. However, the multipoles $E_{2-}$ and $M_{2-}$,
in which this baryon resonance is ``resonant", are not the important
multipoles contributing to the cross-section. It is the $E_{0+}$
multipole, to which $N^*$(2080) contributes as a ``background ",  providing
the bulk of the strength of cross-section. The shape of the cross-section
of Fig.2 is influenced by the rise of the cross-section, as $W$ increases from
the $\eta^{\prime}$ threshold and  the $N^*$(2080) peak is reached. It then
falls, as the strong form factors for the $\eta^{\prime}NR$ and $\eta^{
\prime}NV$ vertices (R, nucleon resonances, V, vector mesons) fall.

Table I shows the contributions\cite{ref18}
 to the real part of the dominant
amplitudes coming from various exchanges.
{\it Note the effective
dominance of the $N^*$(2080)}. Further examination of this reveals the
importance of both the s-channel pole and the non-pole contributions involving
the $N^*$(2080). This is due to the fact that the effective parameter
$\Gamma_{0}\chi_{\lambda}$ controlling this
are predicted\cite{ref1} to be $95$ and $-113$ for
$\lambda =1/2$ and $3/2$ respectively,
for $N^*$(2080), by far the largest. This
parameter is only $-42$ for $N^*$(2090), the next important resonance
excitation\cite{ref1,ref2}.

To summarize,  we are seeing a promising aspect of the
eta prime photoproduction process (1) in being  valuable to explore
the property of $N^*(2080)$, one of the many resonances
around 2 GeV predicted in the quark shell model\cite{ref1,ref3}.
Not much is experimentally known about the electromagnetic and
strong properties of this resonance. Our work shows that the process (1)
would probe these properties. The study of eta-prime photoproduction is
 a nice prospect  for the CEBAF-type facilities.  It
should allow more precise tests of the quark model at this interesting
 region of $W\approx 2 GeV$.
Strong  form factors, about which we know little and want to know more,
are important to reproduce the
experimental data. We have investigated each multipole contribution and found
that the main contribution comes from the $E_{0+}$ multipole. The
surprising fact is that it is the ``background" contribution of a
D13 resonance, $N^* (2080)$, that dominates this multipole, and the
cross-section. This almost mimics a classic resonance behavior,
reminding us again of the difficulty of distinguishing
between a resonance and a non-resonant background\cite{ref19}.
 Due to poor data, we cannot yet give a very  precise estimate of
the properties of $N^* (2080)$. The next theoretical step is to
take into account the unitarity effects in the process, which
requires extensive studies of various decay channels involved in the
reaction (1). This would come with new
experimental initiatives at CEBAF\cite{ref5}.

We are  grateful to R. M. Davidson, J. Napolitano,
 B. G. Ritchie and P. Stoler   for many helpful discussions.
 The research at Rensselaer has been supported by the
U. S. Department of Energy; that   at SAL has been supported
by the Natural Sciences and Engineering Research Council of Canada.



\newpage

Fig. 1: { Calculated total eta prime photoproduction
cross-section using the D13(2080) resonance only.
 The solid line is the D13(2080) contribution
with the form factor given by Eq.(15), the
dashed line is  without the  form factor.}

Fig. 2: {Calculated total eta prime photoproduction cross
section in our effective Lagrangian approach, compared with the experimental
 data\protect\cite{ref4}, stars from
Erbe {\it et al.}\protect\cite{ref4} and
squares from Wolf and S\"{o}ding\protect\cite{ref4}. The parameter
$g_{\eta^{\prime}}$ is taken as 1.9. The solid line is our full calculation,
the dashed line is the  total cross-section obtained
 without the D13(2080)
contribution.}

\newpage

\begin{table}
\caption{Contributions to the real part of the $E_{0^+}$ multipole, in
units of $10^{-3}/m_{\pi+}$, for the $\eta^{\prime}$
photoproduction at the  threshold.
The resonances that are reported in PDG-94 are indicated by $N^*$.
Additional resonances, predicted in the quark model
calculation\protect\cite{ref1}
and not observed yet, are indicated by $R$.}

\label{table12}
\begin{center}
\begin{tabular}{lc}
&\\
Contributions      &   $E_{0+}$ \\[1em] \hline
& \\[1em]
Nucleon Born terms &$-0.55$  \\[1em]
$\rho+\omega$      &$0.89$  \\[1em]
$R(2030)[S11]$      &$0.13$ \\[1em]
$R(2055)[D13]$      &$0.03$ \\[1em]
$N^{*}(2080)[D13]$      &$1.15$ \\[1em]
$N^{*}(2090)[S11]$    &$-0.09$ \\[1em]
$R(2095)[D13]$      &$0.01$  \\[1em]
Total              &$1.56$  \\[1em]
\end{tabular}
\end{center}
\end{table}


\newpage

\begin{thebibliography}{999}

\bibitem{ref1} S. Capstick, {\it Phys. Rev.} {\bf D46}  2864 (1992);
S. Capstick and W. Roberts,{\it Phys. Rev.} {\bf D49}  4570 (1994).
\bibitem{ref2}
For a survey, see L. Montanet {\it et al}. (PDG-94), {\it Phys. Rev.} {\bf
D50},
1173 (1994).
\bibitem{ref3}
N. Isgur, in {\it Proceeding of the CEBAF/SURA 1984 Summer Workshop},
Newport News, Virginia, 1984, F. Gross and R. R. Whitney, eds.,
CEBAF (Newport News, VA, 1984). J. Napolitano, private communication (1994).

\bibitem{ref4}
The Aachen-Berlin-Bonn-Hamburg-Heidelberg-M\"{u}nchen collaboration,
 R. Erbe {\it et al.}, {\it  Phys Rev.} {\bf 175}  1669 (1968).
G. Wolf and P. S\"{o}ding, {\it Electromagnetic Interactions of Hadrons},
{\bf Vol.2}, A. Donnachie and G. Shaw, eds., Plenum (New York,1978).
\bibitem{ref5}
B. G. Ritchie {\it et al}., {\it CEBAF research Proposal}, (October 1, 1991),
and  private communication (1994).
\bibitem{ref6}
F. E. Close, {\it An Introduction to Quarks and Partons}, Academic Press
 (New York, 1979); F. Lenz, {\it Nucl. Phys.} {\bf B279}, 119 (1987).
F. Iachello, N. C. Mukhopadhyay and L. Zhang, {\it Phys. Rev.} {\bf D44},
898 (1991).
\bibitem{ref7}
R. M. Baltrusaitis {\it et al}., {\it Phys. Rev.} {\bf D32}, 2883 (1985).

\bibitem{ref8}
M. Gell-Mann, R. J. Oakes and B. Renner, {\it Phys. Rev.} {\bf 175},
2195 (1968). A. Manohar and H. Georgi, {\it Nucl. Phys.} {\bf B234}, 189
(1984).
U. -G. Mei{\ss}ner, {\it Progr. Theoret. Phys.} {\bf 56}, 903 (1993).
K. Maltman and T. Goldman, {\it Nucl. Phys.} {\bf A572}, 682 (1994).

\bibitem{ref9}
S. Weinberg, {\it Phys. Rev.} {\bf D11}, 3583 (1975); G.'t. Hooft,
{\it Phys. Rev. Lett.} {\bf37}, 8 (1976).  For a brief review,
see S. Coleman, {\it Aspects of Symmetry}, Cambridge Univ. Press
(Cambridge, 1988), p. 307.
\bibitem{ref10}
M. Benmerrouche and N. C. Mukhopadhyay, {\it Phys. Rev. Lett.}
{\bf 67}, 1070 (1991).
M. Benmerrouche, N. C. Mukhopadhyay and J.-F. Zhang, {\it Phys. Rev.}
{\bf D}, in press, (1995).
\bibitem{ref11}
F. Gross, J. W. Van Orden and  K. Holinde, {\it Phys. Rev.} {\bf C41},
R1909 (1990).
\bibitem{ref12}
R. Koniuk and N. Isgur, {\it Phys. Rev.} {\bf D21}, 1868 (1980).
\bibitem{ref13}
M. G. Olsson and E. T. Osypowski, {\it Phys. Rev.} {\bf D17},
174 (1978).
\bibitem{ref14} See, for example,
R. L. Walker, {\it Phys. Rev.} {\bf 182},1729 (1969); J. -M. Laget,
{\it Phys. Report} {\bf 69}, 1 (1981) and private communication (1994).
\bibitem{ref15}
See, for a general discussion on the three-point function, A.O.G. K\"{a}llen
in {\it Dispersion Relations and Elementary Particles}, C. De Witt and
R. Omnes, eds., Hermann (Paris, 1960), p. 418.  For a recent discussion,
see H. Garcilazo and E. Moya de Guerra, {\it Nucl. Phys.} {\bf A562},521
(1993).
\bibitem{ref16}
We are thankful to Dr. R. Davidson for valuable discussions on this point.
\bibitem{ref17}
G. E. Brown, in {\it Excited Baryons}, 1988, G. Adams, N. C.
Mukhopadhyay and P. Stoler, eds., World Scientific (Teaneck, NJ  1989),
p. 17.
\bibitem{ref18}
We use  $g_{\eta^{\prime}}=1.9$
for the nucleon Born terms, and the parameters $\alpha$,
$\beta$ and $\delta$, for the spin-$3/2$ resonances are chosen to be
$-1$, $0.3$ and $-0.8$ for numerical illustration and optimal description
of the data in Fig.2. The parameters $\alpha$, $\beta$ and $\delta$ are
related to $Z$(Eq. (10)) and two other similar ones for the
electromagnetic vertices\protect\cite{ref10}.
\bibitem{ref19}
R. P. Feynman, {\it Photon-Hadron Interactions},
 Benjamin (Reading, MA, 1972). L. Fonda, G. C. Ghirardi
and G. L. Shaw, {\it Phys. Rev.} {\bf D8}, 353 (1973).

\end{thebibliography}
\end{document}